\documentclass[journal=jpclcd,manuscript=letter]{achemso}

\usepackage[version=3]{mhchem} 

\author{Yu Jin}
\affiliation{Department of Chemistry, University of Chicago, Chicago, Illinois 60637, United States}
\altaffiliation{These two authors contributed equally.}
\author{Mariami Rusishvili}
\affiliation{Pritzker School of Molecular Engineering, University of Chicago, Chicago, Illinois 60637, United States}
\altaffiliation{These two authors contributed equally.}
\author{Marco Govoni}
\email{mgovoni@unimore.it}
\affiliation{Department of Physics, Computer Science, and Mathematics, University of Modena and Reggio Emilia, Modena, 41125, Italy}
\alsoaffiliation{Pritzker School of Molecular Engineering, University of Chicago, Chicago, Illinois 60637, United States}
\alsoaffiliation{Materials Science Division, Argonne National Laboratory, Lemont, Illinois 60439, United States}
\author{Giulia Galli}
\affiliation{Pritzker School of Molecular Engineering, University of Chicago, Chicago, Illinois 60637, United States}
\alsoaffiliation{Department of Chemistry, University of Chicago, Chicago, Illinois 60637, United States}
\alsoaffiliation{Materials Science Division, Argonne National Laboratory, Lemont, Illinois 60439, United States}
\email{gagalli@uchicago.edu}

\title{Self-Trapped Excitons in Metal-Halide Perovskites Investigated by Time-Dependent Density Functional Theory}

\abbreviations{IR,NMR,UV}

\usepackage[version=3]{mhchem} 
\usepackage{xcolor}
\usepackage{booktabs}

\begin{document}

\begin{tocentry}
\includegraphics[width=1\textwidth]{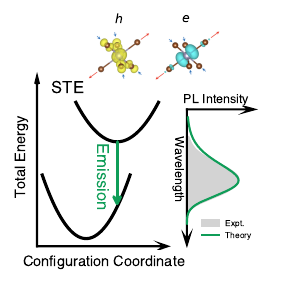}
\end{tocentry}

\begin{abstract}
We present a theoretical study on the formation of self-trapped excitons (STEs) and the associated broadband emission in metal-halide perovskites Cs$_4$SnBr$_6$ and Cs$_2$AgInCl$_6$, using time-dependent density functional theory (TDDFT) with the dielectric-dependent hybrid (DDH) functional. Our approach allows for an accurate description of the excitonic effect and geometry relaxation in the electronic excited states and yields optical gap, STE emission energy, and emission spectra in reasonable agreement with experiments. We point out the significance of considering geometry relaxations in the electronic excited state by showing that the exciton-phonon coupling computed in the ground-state atomic geometry is insufficient to describe the physical properties of STEs. Overall, we find that TDDFT with the DDH hybrid functional is a suitable approach for the study of the formation of STEs in perovskite and provides insights for designing metal-halide perovskites with tailored emission properties. 
\end{abstract}



In recent years, metal-halide perovskites have attracted considerable attention as efficient light emitters~\cite{stranks2015metal,zhang2018all,jena2019halide,liu2021metal,han2021low}. Central to the understanding and design of these luminescent materials is the concept of self-trapped exciton (STE), where an exciton formed upon light absorption becomes trapped into a local potential well caused by a deformation of the crystal lattice induced by the exciton itself. The STE recombination usually results in a broadband light emission with a significant Stokes shift~\cite{cortecchia2019white,du2019microscopic,li2021recent,guo2022light}. First-principles calculations have played an essential role in clarifying the physical origin of the formation of STEs and the broadband emission in several metal-halide perovskite materials, through the investigation of their electronic structure and atomic geometries in the electronic ground state (GS) and excited state (ES)~\cite{wang2019atomistic,du2020first,sun2023toward}. However, controversies remain in comparing results obtained at different levels of theory and, in some cases, also in comparing theory with experiments since measurements are often performed on samples that are not completely defect-free.

Theoretical and computational studies of STEs require the ability to model neutral electronic ESs in periodic systems using large supercells containing hundreds of atoms, given the complex geometry of most halide perovskites. The state-of-the-art approach to the problem would, in principle, be Green's function-based $GW$ method combined with the solution of the Bethe-Salpeter equation ($GW$-BSE)~\cite{onida2002electronic}. However, the unfavorable computational scaling with supercell size and, most importantly, the lack of numerically efficient methods to evaluate nuclear gradients with $GW$-BSE still hinders the applicability of the method to model STEs in metal-halide perovskites~\cite{ismail2003excited,villalobos2023lagrangian}. Therefore, many calculations of STEs in these systems use approaches based on Kohn-Sham (KS) density functional theory (DFT), for example, the constrained-occupation DFT, also called  $\Delta$SCF, where non-Aufbau occupations of KS orbitals are used~\cite{han2018unraveling,li2019origin,du2019microscopic,lian2020photophysics,xu2021toward,jiang2022first,xiong2023first,zhang2023revealing,zhang2023manipulating,liu2023site}, or the restricted open-shell Kohn–Sham (ROKS) method, which is similar to $\Delta$SCF but uses spin-restricted KS orbitals~\cite{luo2018efficient,wang2019atomistic}. With the $\Delta$SCF and the ROKS approaches, one can compute analytical forces acting on nuclei at a computational cost comparable to that of GS DFT calculations. However, these methods are not rigorously derived within a many-body framework and yield results that sometimes need to be further adjusted to compare favorably to experiments~\cite{luo2018efficient,wang2019atomistic}. Moreover, the difficulty in setting the occupation for (near-)degenerate KS orbitals and in converging the calculations has prompted the search for alternative methods.

A method widely used in the chemistry community to obtain neutral excitations and to study the photophysics of molecules is time-dependent DFT (TDDFT)~\cite{runge1984density}. When using hybrid functionals, TDDFT can also capture excitonic effects in solids~\cite{sun2020low}. However, its application in periodic calculations for solids has been limited, mainly due to the lack of efficient implementations for evaluating analytical forces acting on nuclei in the ESs. Recently, by combining a series of algorithms, we obtained an efficient implementation of TDDFT and its analytical nuclear forces in periodic boundary conditions, now available in the open-source code WEST~\cite{jin2023excited}. The controlled numerical approximations used in our implementation, together with an efficient parallelization scheme on both CPU and GPU architectures, allowed for the study of the energy and ES geometry relaxation of point defects in semiconductors and insulators with thousands of electrons~\cite{jin2022vibrationally,verma2023optical}. 

In this letter, we report a hybrid TDDFT study of the electronic structure of two metal-halide perovskites, Cs$_4$SnBr$_6$ and Cs$_2$AgInCl$_6$, focusing on the formation of STEs and the associated broadband emission of these perovskites. These perovskites have shown promise as candidates for efficient light-emitting materials~\cite{luo2018efficient,benin2018highly}, with the former an example of the zero-dimensional (0D) perovskite and the latter a three-dimensional (3D) perovskite. We compare our results with those obtained with $\Delta$SCF calculations and with experiments, with the aim of identifying a robust level of theory to describe STEs in these materials and of assessing the effect of several theoretical approximations.


%
%
In our TDDFT calculations, we used the dielectric-dependent hybrid (DDH) functional~\cite{skone2014ddh,skone2016ddh}, where the fraction of the exact exchange is determined by the inverse of the high-frequency dielectric constant $\epsilon_\infty$ of the solid,  which was computed self-consistently using calculations in finite-field and the Qbox Code~\cite{gygi2008architecture}, resulting in 3.05 for Cs$_4$SnBr$_6$ and 3.85 for Cs$_2$AgInCl$_6$. These values of $\epsilon_\infty$ yield a faction of exact exchange of 0.33 and 0.26, respectively. The DDH functional has been shown to improve the description of the electronic structure of a broad range of systems~\cite{skone2014ddh,skone2016ddh,seo2017designing,gaiduk2016photoelectron,gaiduk2018first,gerosa2018role,pham2017electronic,jin2021pl}, over the results obtained with semi-local functionals. Given the inclusion of screening effects in the Coulomb interaction between the electron and the hole in the excited state~\cite{sun2020low,tal2020accurate,dong2021machine}, we expect excitonic effects to be described accurately as well by the DDH functional.

\begin{figure}
    \centering
    \includegraphics[width=16cm]{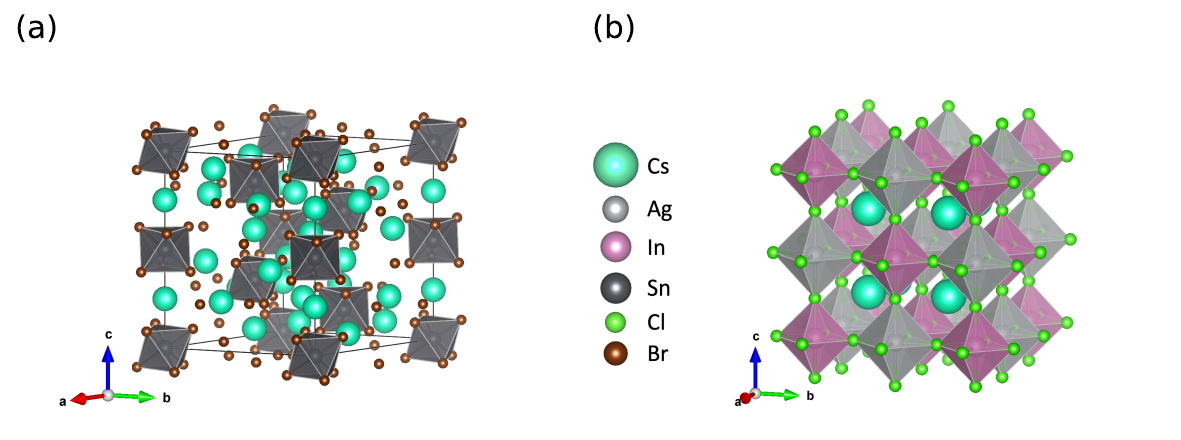}
    \caption{Ball-and-stick model of (a) primitive unit cell of Cs$_4$SnBr$_6$ and (b) conventional unit cell of Cs$_2$InAgCl$_6$. The structure of Cs$_4$SnBr$_6$ belongs to the $R\overline{3}c$ space group, and the structure of Cs$_2$InAgCl$_6$ belongs to the $Fm\overline{3}m$ space group.}
    \label{fig:structure}
\end{figure}

We first computed the fundamental and the optical gaps of Cs$_4$SnBr$_6$ and Cs$_2$AgInCl$_6$ by carrying out DFT and TDDFT calculations, and the results are summarized in Table~\ref{tab:gap} together with theoretical and experimental results reported in the literature. The DFT calculations were performed using the Quantum ESPRESSO code~\cite{giannozzi2020qe,carnimeo2023quantum} and the SG15 optimized norm-conserving Vanderbilt (ONCV) pseudopotentials~\cite{hamann2013optimized,schlipf2015optimization} with the following valence electron configurations: Cs $[4\mathrm{s}^25\mathrm{p}^66\mathrm{s}^1]$, Ag $[4\mathrm{s}^24\mathrm{p}^65\mathrm{s}^14\mathrm{d}^{10}]$, In $[5\mathrm{s}^24\mathrm{d}^{10}5\mathrm{p}^1]$, Cl $[3\mathrm{s}^23\mathrm{p}^5]$, Sn $[5\mathrm{s}^24\mathrm{d}^{10}5\mathrm{p}^2]$, and Br $[4\mathrm{s}^24\mathrm{p}^5]$. A kinetic energy cutoff of 40 Ry was used for the plane-wave basis set, and the Brillouin zone was sampled with the $\Gamma$ point. The TDDFT calculations were carried out using the WEST code~\cite{govoni2015west,yu2022west,jin2023excited} under the Tamm-Dancoff approximation. The analytical forces acting on nuclei were computed using the extended Lagrangian approach by Hutter~\cite{hutter2003excited} implemented within a plane-wave approach~\cite{jin2023excited}.

The solid Cs$_4$SnBr$_6$ is a 0D perovskite, where the [SnBr$_6$] octahedra are not directly connected. Both the valence and conduction bands of Cs$_4$SnBr$_6$ exhibit a weak dispersion~\cite{shi2019impact}; hence, we performed DFT and TDDFT calculations in the primitive unit cell containing 66 atoms. The fundamental gap computed using the DDH functional is larger than those computed using the PBE0 functional, due to the higher fraction of exact exchange (0.33 in DDH, compared to 0.25 in PBE0). The optical gap computed using TDDFT with the DDH functional is also higher than those computed using the $\Delta$SCF approach with the PBE0 functional. The TDDFT (DDH) result overestimates the experimental optical gap reported in literature~\cite{benin2018highly,zhang2019room} by 0.4 eV. This discrepancy may have several origins. From a theoretical standpoint, it might stem from the use of a global dielectric constant in the DDH functional, which may underestimate the screening effects on the valence band maximum (VBM) localized on the [SnBr$_6$] octahedra~\cite{zheng2019dielectric}. The use of a hybrid functional that includes the spatial variation of the dielectric screening~\cite{zhan2023nonempirical} may improve the description of the screening effects on the VBM and will be considered in future works. Additionally, our calculations have been performed at $T=0$ while experiments are performed at finite $T$, and the measured gaps naturally include the dependence on temperature due to exciton-phonon coupling~\cite{alvertis2023phonon}, which may well be substantial. It should be noted that the closer alignment of the optical gap calculated using the PBE0 level of theory with experiments is fortuitous as the screening of the PBE0 functional, corresponding to a dielectric constant ($\epsilon_\infty$) of 4, does not match the system's dielectric constant of 3.05. Note also that the two different PBE0 calculations reported in the literature yield markedly different exciton binding energies at $T=0$ (0.6 and 1.25 eV). Our calculations yield 0.86 eV at $T=0$, and its dependence on $T$ remains to be explored.

\begin{table}[]
    \centering
    \begin{tabular}{cccc}
    \hline\hline
    & & Cs$_4$SnBr$_6$ & Cs$_2$AgInCl$_6$ \\
    \hline
    $E_{\mathrm{gap}}$ (eV) & DDH & 5.15 & 3.19 \\
    & PBE0 & 4.5\textsuperscript{\emph{a}}, 5.01\textsuperscript{\emph{b}}& 2.9\textsuperscript{\emph{e}} \\
    & $GW$ & & 3.27\textsuperscript{\emph{b}} \\
    $E_{\mathrm{opt}}$ (eV) & TDDFT (DDH) & 4.29 & 3.00 \\
    & $\Delta$SCF (PBE0) & 3.90\textsuperscript{\emph{a}}, 3.76\textsuperscript{\emph{b}}& \\
    & $GW$-BSE & & 3.02\textsuperscript{\emph{f}}, 3.22\textsuperscript{\emph{b}} \\
    & Expt. & 3.65\textsuperscript{\emph{c}}, 3.87\textsuperscript{\emph{d}} & 3.3\textsuperscript{\emph{e}} \\
    \hline\hline
    \vspace{0mm}
    \end{tabular}
    
    \textsuperscript{\emph{a}} Ref.~\citenum{shi2019impact}. 
    \textsuperscript{\emph{b}} The $GW$-BSE calculations were performed on top of PBE ground state calculations, as reported in Ref.~\citenum{wang2019atomistic}.     \textsuperscript{\emph{c}} Ref.~\citenum{benin2018highly}. \textsuperscript{\emph{d}} Ref.~\citenum{zhang2019room}.
    \textsuperscript{\emph{e}} Ref.~\citenum{volonakis2017cs2inagcl6}.
    \textsuperscript{\emph{f}} The $GW$-BSE calculations were performed on top of PBE ground state calculations, as reported in Ref.~\citenum{luo2018efficient}.
    \caption{Electronic and optical properties of inorganic halide perovskites, obtained at different levels of theory (see text) and experiments. $E_{\mathrm{gap}}$ and $E_{\mathrm{opt}}$ are the fundamental gap and the optical gap, respectively.}
    \label{tab:gap}
\end{table}

In contrast to the previous case, Cs$_2$AgInCl$_6$ is a three-dimensional (3D) perovskite with all [AgCl$_6$] and [InCl$_6$] octahedra connected by vertices, and the valence and conduction bands are more delocalized~\cite{volonakis2017cs2inagcl6,luo2018efficient,wang2019atomistic} than those of Cs$_4$SnBr$_6$. Therefore, we performed DFT and TDDFT calculations in supercells of different sizes, ranging from 40 to 1080 atoms, and checked the convergence of both the fundamental and optical gaps as a function of size (see Figure~S1 of the supporting information (SI)). The fundamental gap is almost constant as a function of the supercell size, resulting in a value of 3.19 eV in the dilute limit, which agrees well with previous $GW$ results~\cite{wang2019atomistic,luo2018efficient}. The optical gap has a linear dependence on $1/N_{\mathrm{atom}}^{1/3}$, which arises from the electron-hole interaction within the exciton. We extrapolated to the dilute limit by considering an exciton radius of 10.4 \AA\, estimated from the Wannier exciton model using the effective electron and hole masses and dielectric screening~\cite{biega2023chemical}, and we obtained an optical gap of 3.00 eV, in agreement with the experimental value of 3.3 eV~\cite{volonakis2017cs2inagcl6} and $GW$-BSE results of 3.02 -- 3.22 eV~\cite{luo2018efficient,wang2019atomistic}. The agreement with the experiment may slightly worsen if we considered temperature effects~\cite{ha2021quasiparticle}; however, we note that for this system, the exciton binding energy ($\sim$ 0.2 eV) is much smaller than for the 0D perovskite. Hence, the absolute value of the temperature dependence of the exciton-phonon renormalization is not expected to be substantial.

We note that we also attempted $\Delta$SCF calculations of the optical gap of the 0D and 3D perovskites. However, the difficulty in setting occupation numbers for (near-)degenerate KS orbitals prevented us from obtaining a converged optical gap.

\begin{figure}
    \centering
    \includegraphics[width=16cm]{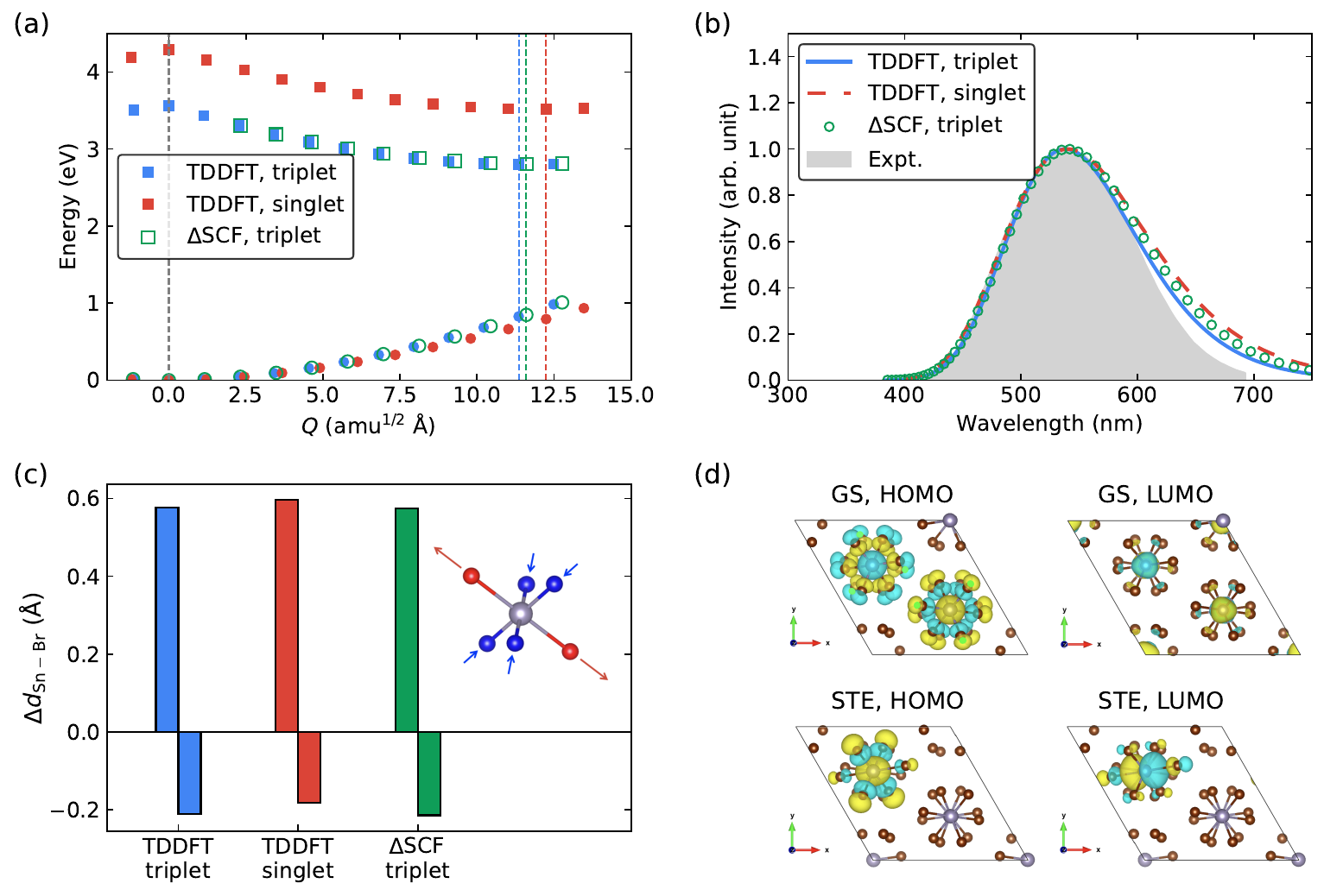}
    \caption{Self-trapped exciton (STE) in Cs$_4$SnBr$_6$. (a) Configuration coordinate diagrams of the STE calculated using the TDDFT and the $\Delta$SCF approach. The squares represent the excited states (ESs), while the circles denote the ground state (GS). The vertical dashed lines denote the atomic geometry of the GS (gray) and the TDDFT triplet ES (blue), the TDDFT singlet ES (red), and the $\Delta$SCF triplet ES (green). The $\Delta$SCF calculations encountered convergence issues around the GS atomic geometry. (b) Computed emission line shapes at 300 K compared with the experimental spectrum from Ref.~\citenum{benin2018highly}. The computed emission line shapes were shifted to align the peak position to the experimental one. (c) Change of the Sn$-$Br bond lengths due to the formation of the STE at a single [SnBr$_6$] octahedron computed using TDDFT and the $\Delta$SCF approach. Two Sn$-$Br bonds are elongated while four Sn$-$Br bonds are contracted in the [SnBr$_6$] octahedron, as shown in the inset. (d) Kohn-Sham orbital density, $|\varphi(\mathbf{r})|^2$, for the highest occupied molecular orbital (HOMO) and the lowest unoccupied molecular orbital (LUMO) obtained from GS DFT calculations at the atomic geometry of the GS and the STE.}
    \label{fig:cs4snbr6_ste}
\end{figure}

%
%
We now turn to the study of the formation of STEs and the associated broadband emission, starting with the Cs$_4$SnBr$_6$ perovskite. We optimized the atomic geometry in the electronic singlet and triplet ESs by performing TDDFT calculations using the DDH functional. For comparison, we also performed $\Delta$SCF calculations for the triplet ES with the same functional. We also attempted $\Delta$SCF calculations for the mixed-spin ES but encountered numerical convergence issues that precluded the completion of the calculation. The resulting STE-related quantities, including the STE emission energy, the self-trapping energy, defined as the energy difference between the free exciton and the STE, and the lattice deformation energy, defined as the energy change in the GS potential energy surface (PES) due to the formation of the STE, are summarized in Table~\ref{tab:hrf}. We computed the mass-weighted displacements between the atomic geometries of the GS and the STE as
\begin{equation}
    \Delta Q= \left[\sum_{\alpha=1}^{N_{\text{atom}}} \sum_{i=x,y,z} M_{\alpha} \left(R_{\alpha i}^{\mathrm{STE}} - R_{\alpha i}^{\mathrm{GS}}\right)^2\right]^{1/2},
\end{equation}
where $M_{\alpha}$ is the mass of the $\alpha$-th atom, and $R_{\alpha i}^{\mathrm{GS}}$ ($R_{\alpha i}^{\mathrm{STE}}$) is the atomic coordinate of the $\alpha$-th atom in the $i$-th direction of the GS (STE). Our geometry optimization led to an STE geometry with significant Jahn-Teller type distortions at a single [SnBr$_6$] octahedron with two Sn$-$Br bonds elongated and four Sn$-$Br bonds contracted, as shown in Figure~\ref{fig:cs4snbr6_ste}(c). Our TDDFT and $\Delta$SCF calculations yield similar geometry distortions. The local geometry distortion leads to the localization of the frontier KS orbitals, at variance with the orbitals at the GS atomic geometry, as shown in Figure~\ref{fig:cs4snbr6_ste}(d).

\begin{table}[]
    \centering
    \begin{tabular}{ccccccccc}
    \hline\hline
    & $E_{\mathrm{emi}}$ & $E_{\mathrm{st}}$ & $E_{\mathrm{d}}$ & $\Delta Q$ & $\hbar\Omega_{\mathrm{gs}}$ &  $\hbar\Omega_{\mathrm{es}}$ & $S_{\mathrm{gs}}$ & $S_{\mathrm{es}}$ \\
    &(eV) & (eV) & (eV) & (amu$^{0.5}$ \AA) & (meV) & (meV) & & \\
    \hline
    \addlinespace[2mm]
    &\multicolumn{8}{c}{Cs$_4$SnBr$_6$} \\
    $\Delta$SCF (DDH), triplet & 1.96 & & 0.85 & 11.6 & 8.06 & 7.30 & 130 & 118 \\
    TDDFT (DDH), triplet & 1.97 & 0.76 & 0.83 & 11.4 & 8.11 & 7.62 & 125 & 117 \\
    TDDFT (DDH), singlet & 2.72 & 0.77 & 0.79 & 12.2 & 7.62 & 6.47 & 137 & 116 \\
    $\Delta$SCF (PBE0), triplet \textsuperscript{\emph{a}} & 2.27 & 0.72 & 0.77 &  & 6.59 & 6.72 & 117 & 107 \\
    Expt. & 2.30 \textsuperscript{\emph{b}} & & & & & & & \\
    \hline
    \addlinespace[2mm]
    &\multicolumn{8}{c}{Cs$_2$AgInCl$_6$} \\
    $\Delta$SCF (DDH), triplet & 1.28 & & 1.15 & 5.56 & 19.9 & 16.4 & 74 & 61 \\
    TDDFT (DDH), triplet & 1.32 & 0.38 & 1.04 & 5.54 & 18.7 & 16.4 & 69 & 60 \\
    TDDFT (DDH), singlet & 1.21 & 0.42 & 1.13 & 5.68 & 19.4 & 16.3 & 75 & 63 \\
    ROKS (PBE), singlet \textsuperscript{\emph{c}} & 1.82 & 0.53 & 0.67 &  & 18.3 & 17.4 & 37 & 30 \\
    Expt. & 2.04 \textsuperscript{\emph{d}}  & & & & & & & \\
    \hline\hline
    \vspace{0mm}
    \end{tabular}
    
    \textsuperscript{\emph{a}} Results obtained using the $\Delta$SCF approach with the PBE0 functional as reported in Ref.~\citenum{wang2019atomistic}. 
    \textsuperscript{\emph{b}} Ref.~\citenum{benin2018highly}.  \textsuperscript{\emph{c}} Results obtained using the restricted open-shell Kohn-Sham (ROKS) approach with the PBE functional and the scaled Perdew-Zunger self-interaction correction (PZ-SIC) on the unpaired electrons as reported in Ref.~\citenum{luo2018efficient,wang2019atomistic}. The scaling parameter of the Hartree energy was fitted to reproduce the exciton binding energies calculated by the $GW$-BSE approach. The emission energy was adjusted by accounting for the discrepancy between the free-exciton energy derived from the $GW$-BSE calculation and that from the ROKS calculation. \textsuperscript{\emph{d}} Ref.~\citenum{volonakis2017cs2inagcl6} \\
    \caption{Computed properties of self-trapped excitons (STEs) in inorganic halide perovskites. $E_{\mathrm{emi}}$, $E_{\mathrm{st}}$, and $E_{\mathrm{d}}$ are the emission energy, self-trapping energy, lattice deformation energy, respectively. $\Delta Q$ is the mass-weighted displacements between the atomic geometries of the ground state and the STE. $\omega_{\mathrm{gs}}$ ($\omega_{\mathrm{es}}$) is the effective phonon frequency, and $S_{\mathrm{gs}}$ ($S_{\mathrm{es}}$) is the corresponding Huang-Rhys factor for the ground state (excited state). We show results obtained at different levels of theory (see text).}
    \label{tab:hrf}
\end{table}

We then built configuration coordinate diagrams between the atomic geometries of the GS and STE using a linear interpolation method, as shown in Figure~\ref{fig:cs4snbr6_ste}(a). We note that the energy curves of the TDDFT triplet ES and $\Delta$SCF triplet ES are similar, indicating that the $\Delta$SCF approach provides reasonable results for the triplet STE, although we encountered numerical convergence issues in $\Delta$SCF calculations for configurations near the GS atomic geometry. The energy curves of the TDDFT singlet ES and the TDDFT triplet ES have similar shapes but differ by a rigid shift. The singlet STE has an energy of 0.75 eV higher than that of the triplet STE due to the additional exchange interaction between the electron and the hole localized on a single [SnBr$_6$] octahedron. A cusp can be observed in the energy curves of the TDDFT singlet and triplet ESs at the GS atomic geometry. By extrapolating in the negative direction along the configuration coordinate diagram, which corresponds to the contraction of two Sn$-$Br bonds and the elongation of the other four Sn$-$Br bonds, we observed a saddle point, as shown in Figure~S2 of the SI, which is a signature of Jahn-Teller distortions.

The configuration coordinate diagram allows us to compute the emission line shape for the STE, which corresponds to the transition from the STE ES PES to the GS PES, using Fermi's golden rule and the Franck-Condon principle~\cite{wang2019atomistic,jin2021pl}:
\begin{equation}
    L(\hbar \omega, T) \propto (\hbar\omega)^3 \sum_{i,j} P_{e j}(T)\left|\left\langle\chi_{e j} \mid \chi_{g i}\right\rangle\right|^{2} \delta\left(E_{\mathrm{ZPL}}+j\hbar\Omega_e - i\hbar\Omega_g  -\hbar \omega\right),
\end{equation}
where $\hbar\omega$, $T$, and $E_{\mathrm{ZPL}}$ represent the photon energy, the temperature, and the energy of zero-phonon line emission, respectively. We adopted the one-dimensional effective phonon approximation for the nuclear wave functions, where $|\chi_{gi}\rangle$ ($|\chi_{ej}\rangle$) represents the $i$-th ($j$-th) vibrational wave function of the quantum harmonic oscillator with the effective energy of $\hbar\Omega_g$ ($\hbar\Omega_e$). The overlap integrals $\left|\left\langle\chi_{e j} \mid \chi_{g i}\right\rangle\right|^{2}$ can be calculated recursively using $\hbar\Omega_g$, $\hbar\Omega_e$ and $\Delta Q$~\cite{ruhoff1994recursion}. The exciton-phonon coupling strength for the formation of STE can be characterized by the Huang-Rhys factor (HRF), computed as~\cite{jin2021pl}
\begin{equation}
    S_g = \dfrac{\Omega_g \Delta Q^2}{2\hbar}, \quad S_e = \dfrac{\Omega_e \Delta Q^2}{2\hbar},
\end{equation}
for the GS and the ES, respectively; this quantity can be viewed as the average number of phonons emitted during the optical transition.

\begin{figure}
    \centering
    \includegraphics[width=10cm]{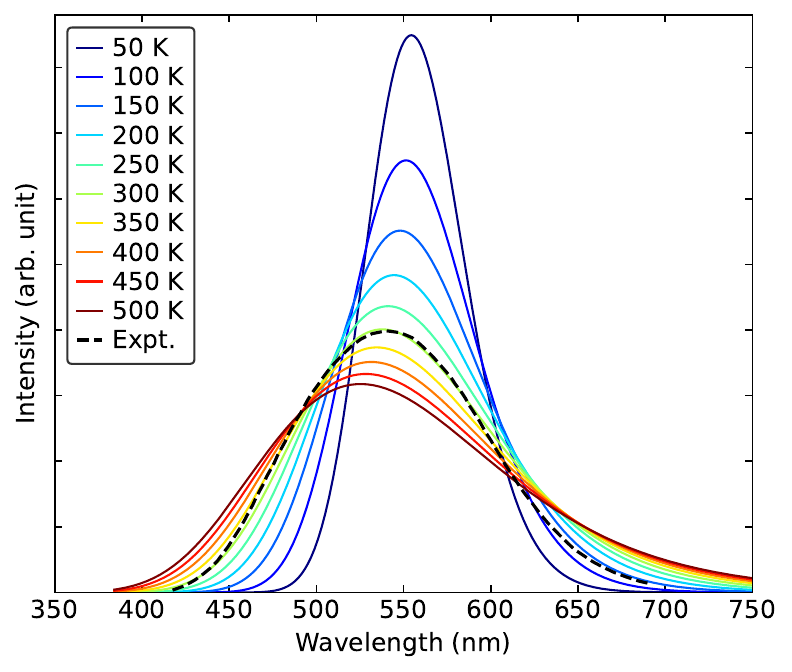}
    \caption{Computed temperature dependent emission line shapes for Cs$_4$SnBr$_6$ compared with the experimental spectrum at 300 K~\cite{benin2018highly}. The computed emission line shapes were shifted to align the position of the peak to the experimental one.}
    \label{fig:cs4snbr6_tdpl}
\end{figure}

The calculated emission line shapes of the STE are shown in Figure~\ref{fig:cs4snbr6_ste}(b), together with the experimental emission line shape. The line shapes computed using parameters from different ES calculations agree well with each other, and they are in reasonable agreement with experiments, except for a slight overestimate of the peak width. The broadband feature of the emission line shapes can be related to the strong coupling between excitons and phonons in the optical transition, reflected by the large HRFs as shown in Table~\ref{tab:hrf}. We also examined the temperature dependence of the emission line shape, as shown in Figure~\ref{fig:cs4snbr6_tdpl}. Interestingly, a similar temperature dependence has been observed experimentally for Cs$_4$PbBr$_6$~\cite{yin2017intrinsic}.

The computed emission line shapes with TDDFT for the singlet and triplet STEs agree with the experiments. However, the calculated emission energy for the singlet STE is 0.42 eV higher than the experiment, while the computed emission energy for the triplet STE is 0.33 eV lower than the experiment. The STE radiative lifetime of 500 ns obtained by time-resolved photoluminescence measurements suggests that the broadband emission might originate from the triplet STE~\cite{benin2018highly,zhang2019room}. We note that the direct calculation of the triplet emission lifetime requires evaluating the spin-orbit coupling strength between singlet and triplet excited states, which is beyond the scope of the current study.

One possible reason for the underestimation of our calculated STE emission energy and the overestimation of the emission peak width compared to the experiment could be from the use of the global dielectric constant in the DDH hybrid functional; such functional is expected to underestimate the screening effect on the distorted [SnBr$_6$] octahedron where the frontier KS orbitals are localized, leading in turn to an overestimated local geometric distortion.

%
%
To investigate the exciton-phonon coupling for the STE formation in Cs$_4$SnBr$_6$, we examined the partial HRFs for each phonon mode. The partial HRF on the $k$-th phonon mode, $S_k$, can be computed as $S_k = \frac{\omega_k\Delta Q_k^2}{2\hbar}$, where $\omega_k$ is the vibrational frequency of the $k$-th phonon mode, and $\Delta Q_k$ is the projection of the mass-weighted displacements on the $k$th phonon mode, which is evaluated using the displacement between the atomic coordinates of the GS and the STE
\begin{equation}
    \Delta Q_{k}^{\mathrm{Dis}} = \sum_{\alpha=1}^{N_{\text{atom}}} \sum_{i=x,y,z} \sqrt{M_\alpha} \left(R_{\alpha i}^{\mathrm{STE}} - R_{\alpha i}^{\mathrm{GS}} \right) e_{k,\alpha i},
\end{equation}
where $\mathbf{e}_k$ is the eigenvector of the $k$-th phonon mode. Using the displaced harmonic oscillator approximation, where the GS and the ES PESs are assumed to have the same shape except for a displacement, $\Delta Q_k$ can be calculated equivalently using nuclear forces,
\begin{equation}
    \Delta Q_k^{\mathrm{Forces}} = \frac{1}{\omega_k^2} \sum_{\alpha=1}^{N_{\mathrm{atom}}} \sum_{i=x, y, z} \frac{F_{\alpha i}^{\mathrm{ES@GS}}}{\sqrt{M_\alpha}} e_{k, \alpha i},
\end{equation}
where $\mathbf{F}^{\mathrm{ES@GS}}$ represents the atomic forces on the ES PES at the GS atomic geometry. Since $\mathbf{F}^{\mathrm{ES@GS}}$ equals the derivative of the excitation energy with respect to the atomic coordinates at the GS atomic geometry, $S_k$ can be regarded as the exciton-phonon coupling strength of the $k$-th phonon mode.

\begin{figure}
    \centering
    \includegraphics[width=10cm]{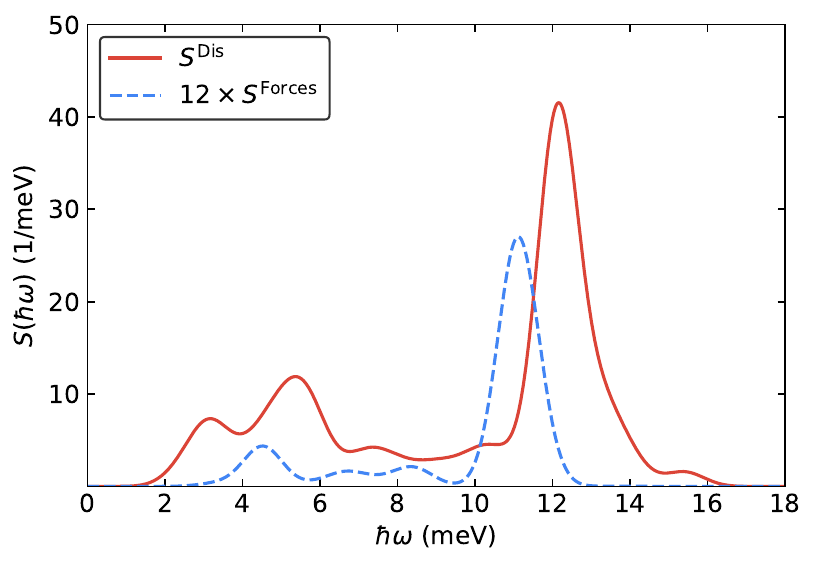}
    \caption{Spectral density of the Huang-Rhys factors (HRFs) as a function of the phonon energy for Cs$_4$SnBr$_6$. The HRFs are computed either using the displacements between the atomic geometries of the ground state (GS) and the STE, denoted as $S^{\mathrm{Dis}}$ (red solid line), or using the excited state forces at the GS atomic geometry, denoted as $S^{\mathrm{Forces}}$ (blue dashed line). $S^{\mathrm{Forces}}$ was multiplied by 12 times for a better visual comparison with $S^{\mathrm{Dis}}$.}
    \label{fig:cs4snbr6_exph}
\end{figure}

We computed the partial HRFs, $S_k$, using the two approaches, and displayed the spectral density, $S(\hbar\omega) = \sum_k S_k \delta(\hbar\omega - \hbar\omega_k)$, in Figure~\ref{fig:cs4snbr6_exph}. The spectral density computed using displacements ($S^{\mathrm{Dis}}$) and using forces ($S^{\mathrm{Forces}}$) are different in terms of both the peak position and peak intensity. The peak at 12 meV for $S^{\text{Dis}}$ is related to the Jahn-Teller type phonon modes, consistent with the Jahn-Teller type of geometry distortion due to the formation of the STE. The peak at 11 meV for $S^{\text{Forces}}$ is related to the bending type phonon modes, and the intensity is more than one magnitude smaller. The drastic difference between $S^{\mathrm{Dis}}$ and $S^{\mathrm{Forces}}$ indicates that the exciton-phonon coupling at the GS atomic geometry is not appropriate to describe the formation of the STE; direct geometry optimization in ES PES is required instead.

\begin{figure}
    \centering
    \includegraphics[width=16cm]{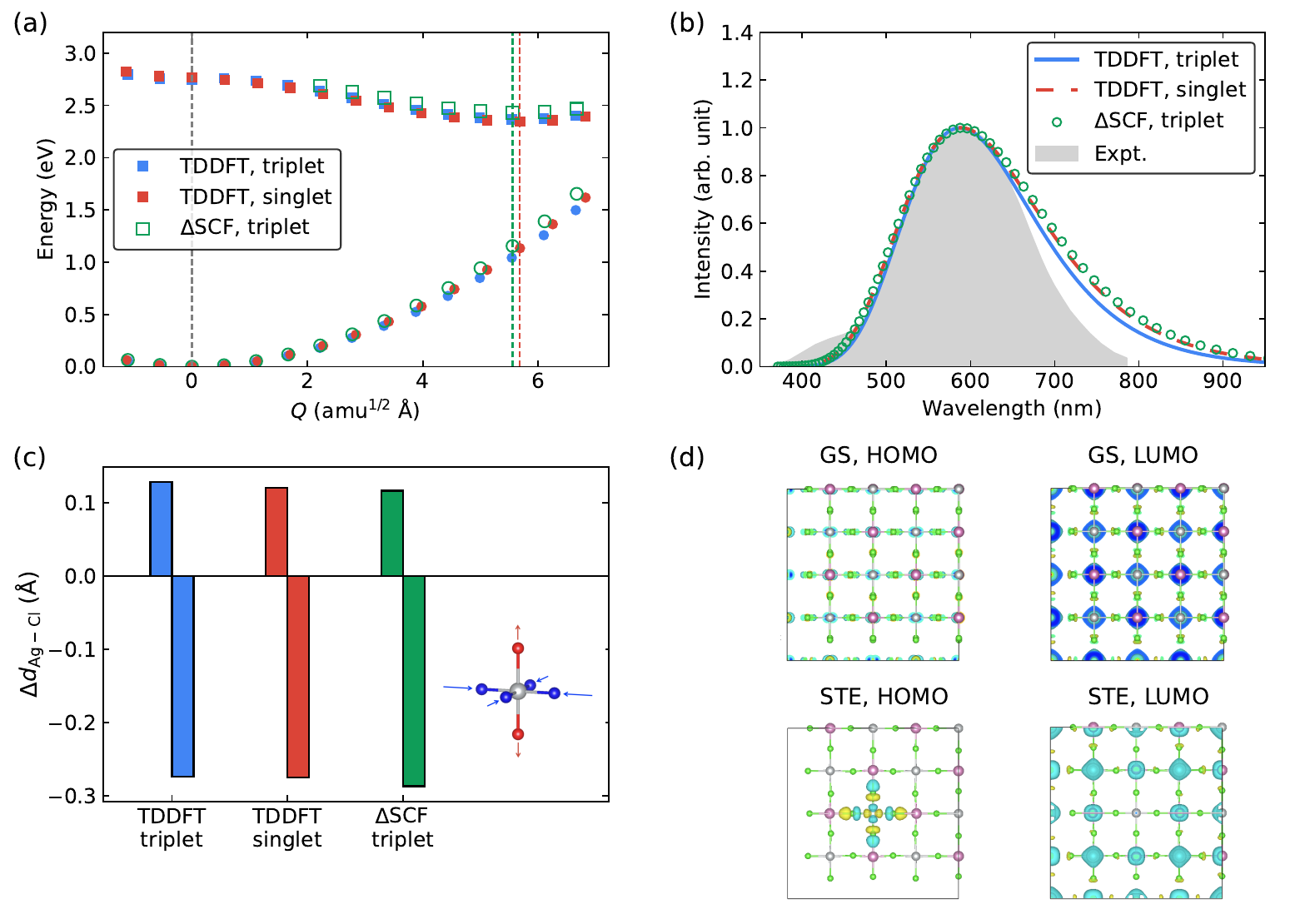}
    \caption{Self-trapped exciton (STE) in Cs$_2$AgInCl$_6$. (a) Configuration coordinate diagrams of the STE calculated using the TDDFT and the $\Delta$SCF approach. The squares represent the excited states (ESs), while the circles denote the ground state (GS). The vertical dashed lines denote the atomic geometry of the GS (gray) and the TDDFT triplet ES (blue), the TDDFT singlet ES (red), and the $\Delta$SCF triplet ES (green). The $\Delta$SCF calculations encountered convergence issues around the GS atomic geometry. (b) Computed emission line shapes at 300 K compared with the experimental spectrum from Ref.~\citenum{luo2018efficient}. The computed emission line shapes were shifted to align the peak position to the experimental one. (c) Change of the Ag$-$Cl bond lengths due to the formation of the STE at a single [AgCl$_6$] octahedron computed using TDDFT and $\Delta$SCF approach. Two Ag$-$Cl bonds are elongated while four Ag$-$Cl bonds are contracted in the [AgCl$_6$] octahedron, as shown in the inset. (d) Kohn-Sham orbital density, $|\varphi(\mathbf{r})|^2$, for the highest occupied molecular orbital (HOMO) and the lowest unoccupied molecular orbital (LUMO) obtained from GS DFT calculations at the atomic geometry of the GS and the STE.}
    \label{fig:cs2agincl6_ste}
\end{figure}

%
%
In a similar fashion, we also studied the formation of the STE in  Cs$_2$AgInCl$_6$. A $(2\times2\times2)$ conventional supercell containing 320 atoms was used to localize the STE. The computed STE emission energy, self-trapping energy, and lattice deformation energy are summarized in Table~\ref{tab:hrf}. The configuration coordinate diagram, the optical emission line shape, and the geometry distortion due to the formation of the STE and the localization of frontier KS orbitals are displayed in Figure~\ref{fig:cs2agincl6_ste}.

The TDDFT geometry optimization in the ES shows a Jahn-Teller type geometry distortion on a single [AgCl$_6$] when the STE is formed, as shown in Figure~\ref{fig:cs2agincl6_ste}(c), which is consistent with the results of previous theoretical studies~\cite{luo2018efficient,wang2019atomistic}. At the STE geometry, the highest occupied molecular orbital (HOMO) is localized on a single  [AgCl$_6$], while the lowest unoccupied molecular orbital (LUMO) is still delocalized and repelled away from the [AgCl$_6$] octahedron, consistent with the dispersion of the VBM and CBM~\cite{volonakis2017cs2inagcl6} found in our calculations. These results are consistent with the fact that the TDDFT singlet and triplet ESs have similar emission energies, as a result of minimizing the exchange interaction between the hole at the localized HOMO and the electron at the delocalized LUMO. This is in contrast to the case for Cs$_4$SnBr$_6$, where the HOMO and LUMO are localized on the same [SnBr$_6$] octahedron. The STE geometry distortions (Figure~\ref{fig:cs2agincl6_ste}(c)), the configuration coordinate diagrams (Figure~\ref{fig:cs2agincl6_ste}(a)), and the optical emission line shapes (Figure~\ref{fig:cs2agincl6_ste}(b)) for the triplet ES computed using TDDFT and $\Delta$SCF are  similar. However, we note that $\Delta$SCF calculations failed to converge near the GS geometry, probably due to the existence of near-degenerate orbitals. In contrast with previous studies that employed the ROKS approach and revealed a significant gradient in the ES PES near the GS geometry~\cite{luo2018efficient}, the ES PES in the vicinity of the GS geometry computed using TDDFT exhibits a relatively flat profile. The transition from the flat PES to the parabolic PES for the ES suggests a transition from the free exciton to the STE, a process that our TDDFT calculation successfully captures.

The exciton emission energy computed using TDDFT and $\Delta$SCF underestimates the experimental emission energy by 0.7 -- 0.8 eV, which is associated with the overestimation of the STE geometry distortion that leads to a broader emission line shape than in experiments, as shown in Figure~\ref{fig:cs2agincl6_ste}(b). The discrepancy between our theoretical results and experiments can be ascribed to several factors. Once more, the use of a global dielectric constant in the DDH functional may be responsible for inaccuracies in the description of the screening effect for the localized HOMO at the STE geometry, thus resulting in a deficient description of the electronic and geometric properties of the STE. In addition, our TDDFT calculations, despite employing a 320-atom supercell, might still be influenced by finite-size effects, which may be particularly relevant due to the delocalized nature of the LUMO in the STE geometry. Employing a finite-size correction of 0.2 eV, the same as that used for the free exciton in the 320-atom supercell (see our extrapolation in Figure~S1), the TDDFT-calculated STE emission energy would become 1.5 eV. Further accounting for a 0.3 eV discrepancy between the calculated and experimental optical gaps, we obtain an STE emission of 1.8 eV. We are not in a position to assess the robustness of these corrections at this time, and we approach the interpretation of the 2.0 eV emission peak to STE with caution. There might be other mechanisms not considered here that come into play, such as the emission from point defects, for example, antisite defects $\mathrm{In}_{\mathrm{Ag}}$~\cite{ha2021origin}, that should be explored in greater detail.

\begin{figure}
    \centering
    \includegraphics[width=10cm]{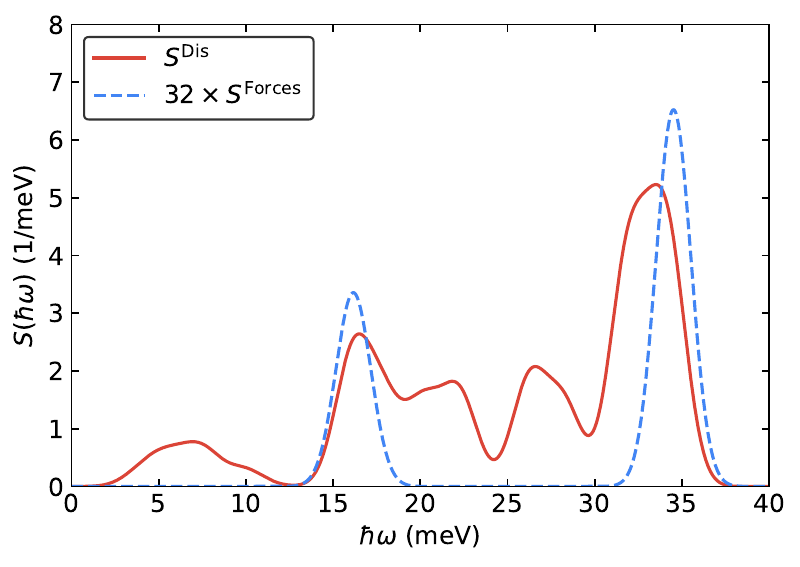}
    \caption{Spectral density of the Huang-Rhys factors (HRFs) as a function of the phonon energy for Cs$_2$AgInCl$_6$. The HRFs are computed either using the displacements between the atomic geometries of the ground state (GS) and the self-trapped exciton (STE), denoted as $S^{\mathrm{Dis}}$ (red solid line), or using the excited state forces at the GS atomic geometry, denoted as $S^{\mathrm{Forces}}$ (blue dashed line). $S^{\mathrm{Forces}}$ was multiplied by 32 times for a better visual comparison with $S^{\mathrm{Dis}}$.}
    \label{fig:cs2agincl6_exph}
\end{figure}

%
%
We also studied the exciton-phonon coupling in Cs$_2$AgInCl$_6$ by computing the partial HRFs using either the displacement between the GS and STE geometries or the atomic forces on the ES PES at the GS geometry, and the resulting spectral densities are displayed in Figure~\ref{fig:cs2agincl6_exph}. Although $S^{\mathrm{Dis}}$ and $S^{\mathrm{Forces}}$ share common peaks at 16 meV and 34 meV, with the former corresponding to the Jahn-Teller type phonon modes and the latter related to the stretching modes, their intensity shows a difference of more than twenty times. Therefore, the exciton-phonon coupling computed at the GS geometry is insufficient for the description of the formation of the STE; the direct ES geometry relaxation is critical.

In summary, we studied the electronic and optical properties of all inorganic halide perovskites Cs$_4$SnBr$_6$ and Cs$_2$AgInCl$_6$, especially the formation of the STE and the associated broadband emission, using TDDFT with the DDH functional. We showed that the approach adopted here allows for an accurate description of the excitonic effects and geometry relaxation in the excited states and yields optical gap, STE emission energy, and optical emission spectra in reasonable agreement with experiments. In addition, TDDFT also allows for the description of the transition from the free exciton to STE. Importantly, we also demonstrated that the exciton-phonon coupling computed at the atomic geometry of the GS is insufficient for the study of STEs and broadband emission; the geometry relaxation in ES PES is required instead. Overall, we found that $\Delta$SCF-based methods, while adequate in some cases, present convergence issues and yield results that are not as accurate as those of TDDFT.

However, some discrepancies between the results obtained here and the experiments remain. In the future, we plan to use hybrid functionals that can describe the spatial variation of the dielectric function~\cite{zheng2019dielectric,zhan2023nonempirical} in (TD)DFT calculations of perovskite systems and to include spin-orbit coupling, which could further improve the agreement with the experiments. Finally, a careful understanding of the influence of defects on measured emission spectra will need to be explored in detail~\cite{zhou2021defect}. 


\begin{acknowledgement}
    We thank Dr. Victor Wen-zhe Yu and Jiawei Zhan for helpful discussions. The theoretical and computational developments of this work were supported by the Computational Materials Science Center Midwest Integrated Center for Computational Materials (MICCoM). MICCoM is part of the Computational Materials Sciences Program funded by the U.S. Department of Energy, Office of Science, Basic Energy Sciences, Materials Sciences, and Engineering Division through the Argonne National Laboratory, under Contract No. DE-AC02-06CH11357. The application to perovskite materials was supported by Center for Hybrid Organic Inorganic Semiconductors for Energy (CHOISE) an Energy Frontier Research Center funded by the Office of Basic Energy Sciences, Office of Science within the U.S. Department of Energy (DOE). Our research used resources of the National Energy Research Scientific Computing Center (NERSC), a DOE Office of Science User Facility supported by the Office of Science of the U.S. Department of Energy under Contract No. DE-AC02-05CH11231 using NERSC award ALCC-ERCAP0025950, and resources of the University of Chicago Research Computing Center.
\end{acknowledgement}

\begin{suppinfo}
    The Supporting Information is available free of charge at \url{http://pubs.acs.org}.

    Extrapolation of the fundamental and optical gaps of Cs$_2$AgInCl$_6$. Configuration coordinate diagram of Cs$_4$SnBr$_6$.
\end{suppinfo}

\bibliography{Main.bib}

\end{document}


\newpage

\begin{figure}
    \centering
    \includegraphics[width=10cm]{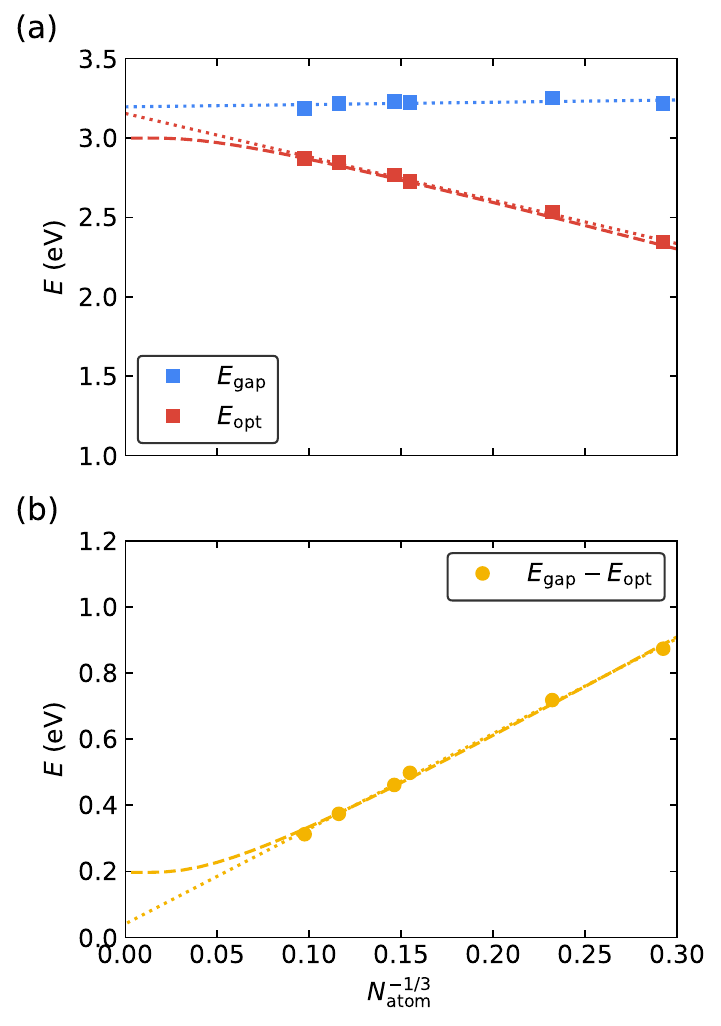}
    \caption{(a) Calculated fundamental gap ($E_{\mathrm{gap}}$) and optical gap ($E_{\mathrm{opt}}$), and (b) exciton binding energy ($E_{\mathrm{gap}} - E_{\mathrm{opt}}$) for Cs$_2$AgInCl$_6$ plotted against the inverse cube root of the number of atoms ($N_{\mathrm{atoms}}^{-1/3}$) in the supercell. The dotted lines represent linear fits of the energies as a function of $N_{\mathrm{atoms}}^{-1/3}$. The dashed lines correspond to non-linear fits incorporating a screening length ($D$) of 38.3 \AA, calculated using $D = D_{\mathrm{H}} \frac{\epsilon_{\infty}}{\mu}$, where $D_{\mathrm{H}} = 1.9$ \AA\ is the hydrogen atom's screening length from DFT calculations~\cite{zhang2020optically}, $\epsilon_\infty = 3.85$ is the dielectric constant of Cs$_2$AgInCl$_6$, and $\mu = 0.191$ is the effective mass of the Wannier-Mott exciton~\cite{biega2023chemical}.
    }
    \label{fig:s1}
\end{figure}

\begin{figure}
    \centering
    \includegraphics[width=10cm]{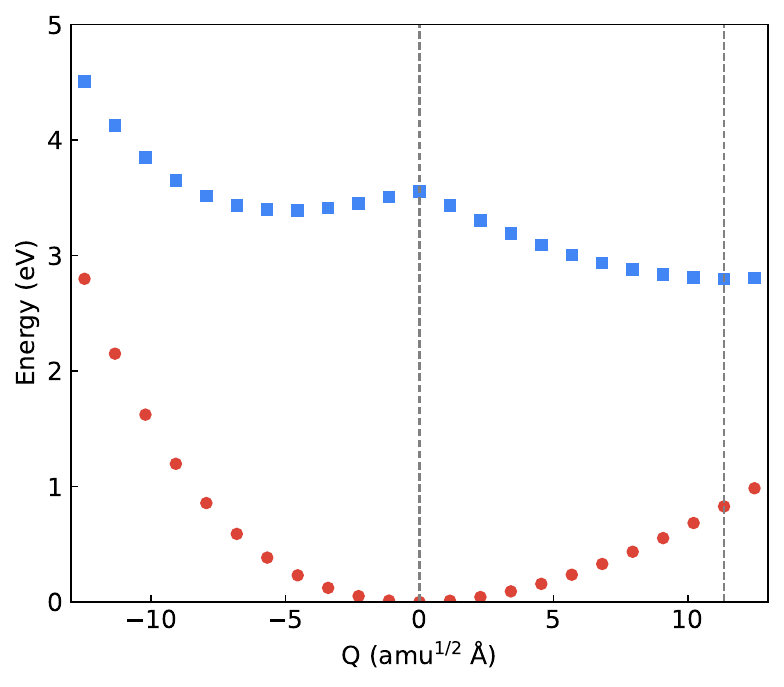}
    \caption{Configuration coordinate diagrams of Cs$_4$SnBr$_6$ computed using the TDDFT approach for the triplet excited state.}
    \label{fig:s2}
\end{figure}



\newpage
\bibliography{SI/SI.bib}